\documentclass[eqsecnum,aps,nofootinbib,preprint,showpacs,preprintnumbers,amsmath,amssymb]{revtex4}

\usepackage{graphicx}% Include figure files
\usepackage{dcolumn}% Align table columns on decimal point
\usepackage{bm}% bold math

\def\be{\begin{equation}}

\def\ee{\end{equation}}

\def\bea{\begin{eqnarray}}

\def\eea{\end{eqnarray}}

\def\ba{\begin{array}}

\def\ea{\end{array}}

 \def\tS{\tilde S}

 \def\hT{\hat T}

\def\ha{\hat a}      

      \def\tb{\tilde b}

\def\hc{\hat c}      

\def\hm{\hat m}      

\def\hr{\hat r}      

\def\hatt{\hat t}

\begin{document}

\title{On Fast Travel through spherically symmetric spacetimes}

\author{Belkis Cabrera Palmer}

\affiliation{
Physics Department, Syracuse University, Syracuse, NY 13244.}

\author{Donald Marolf}

\affiliation{
Physics Department, Syracuse University, Syracuse, NY 13244.}

\date{\today}

%%%%%%%%%%%%%%%%

\begin{abstract}
In a static spacetime, the Killing time can be used to measure the
time required for signals or objects to propagate between two of its orbits.
By further restricting to spherically symmetric cases, one obtains a natural
association between these orbits and timelike lines in Minkowski space.
We prove a simple theorem to the effect that in any spacetime satisfying the weak
energy condition the above signaling time is, in this sense, no faster than that for a corresponding signal
in Minkowski space.
The theorem uses a normalization of Killing time appropriate to an observer at infinity. We
then begin an investigation of certain related but more local questions by studying particular families of spacetimes in detail.   Here we are also interested in restrictions imposed by the dominant energy condition.  Our examples suggest that signaling in spacetimes satisfying this stronger energy condition may be significantly slower than the fastest spacetimes satisfying only the weak energy condition.
\end{abstract}

\maketitle

%%%%%%%%%%%%%%%%%%%%%%%%%%%%%%%%%%%%%%%%%%%%%%%%%%%%%%%%%

\section{Introduction}

\label{intro}

In familiar settings, gravity has a tendency to slow the transfer of information from one place to
another.  The linearized version of this effect is known as the Shapiro time delay  and has been the
subject of many precision tests of general relativity \cite{Will}.  The locally measured speed of light
remains constant, but the curvature of spacetime nevertheless requires a signal travelling between two
``locations'' $x,y$ to take longer than would be required to signal between the corresponding locations
in Minkowski space.

We are interested here in whether this delay can take the form of
an {\it advance}, so that the curved spacetime is in some sense
`faster' than Minkowski space. Several
such senses have been used in the literature and we will introduce
more below.
 This issue has a long history
and allows for interesting speculations.
The question is often asked whether a technologically advanced civilization might alter the spacetime geometry to take a form that is
maximally convenient for their transportation and communication needs
and what bounds exist in principle on their ability to do so.
Another motivation comes from cosmology, as any spacetime of use to
such an advanced civilization might provide for what is in effect a `variable speed of light
cosmology' \cite{vsl1,vsl2,vsl3,vsl4,vsl5,vsl6,vsl7,vsl8} within the context of traditional Einstein-Hilbert gravity.

Of course, the Einstein equations themselves impose no
restrictions on the geometry as they merely
relate curvature to the matter content of the universe.  Thus, any spacetime may be constructed
so long as one includes a sufficiently exotic matter source.
Well known examples of
spacetimes widely considered to be `fast' include those of Alcubierre \cite{Alc} and Krasnikov
\cite{Kras} as well as wormhole solutions (see {\it e.g.} \cite{MTY}).  The
negative mass Schwarzschild solution also allows fast signaling. In fact, all
known solutions which are readily
agreed to be fast contain matter that violates the weak energy condition.  This condition \cite{WaldBook} requires the stress-energy tensor to satisfy
\begin{equation}
\label{WEC} T_{\mu \nu} t^\mu t^\nu \ge 0
\end{equation}
for any timelike vector\footnote{A somewhat weaker condition known as the weak null energy condition is
sometimes used in which the timelike vector $t^\mu$ is replaced by an arbitrary null vector $k^\mu$.
However, we consider only the stronger version based on timelike vectors.} $t^\mu$. In the absence of a
negative cosmological constant, forms of matter violating (\ref{WEC}) are widely believed not to exist,
or at least to be severely limited by fundamental principles.  See {\it e.g.}
\cite{Tipler,Romans1,FW,FR1,Borde} for a summary of the current understanding of the limitations on negative
energy fluxes from quantum field theoretic effects\footnote{Reference \cite{Kras2} provides a somewhat different perspective on
these results.} and \cite{MR} for some discussion of the relationship
between the `negative energies' of stringy orientifolds and the weak energy condition.

There are substantial theorems to the effect that fast travel is not possible without violating this
energy condition. Such theorems include the results of Hawking \cite{Chron} on the formation of closed
timelike curves and those of Olum \cite{Olum}, Visser, Bassett, and Liberati \cite{VBL}, and Gao and Wald
\cite{GW} which relate more directly to `fast travel'. These theorems can be quite powerful and each
involves a somewhat different concept of `fast travel'. Visser, Bassett, and Liberati \cite{VBL} focus on
a perturbative description about flat space. Gao and Wald essentially discuss signaling between points in
various asymptotic regions. Olum takes a different approach and derives a rather more abstract theorem
showing that without violations of (\ref{WEC}), a certain de-focusing property cannot arise.  This
property is expected to be related to localized regions of fast travel, and in particular the idea that
there is a `fastest' path to follow.

However, only the perturbative results of \cite{VBL}
provide actual bounds on signaling times between locations
within the interior of a spacetime. There
is thus a sizable gap in the literature in terms of what one
might call `concrete' results referring to
the interior of a spacetime in the non-perturbative context.
As discussed in \cite{VBL}, the basic difficulty is to find a setting in
which one may ask a well-defined question.  One would like to ask whether one spacetime is `faster' than
another, but this would require some way to identify standard `locations' in the two
spacetimes between which one wishes to travel.  The diffeomorphism invariance of general relativity is
well-known to make such notions extremely difficult to define.

We begin to fill this gap below by using the restricted context of spherically symmetric static
spacetimes to ask well-defined questions. In section \ref{bound}, we derive a non-perturbative version of
\cite{VBL} within this context.  Roughly speaking, it states that in terms of the time $T$
measured by an
observer at infinity, a signal between any two orbits $x,y$ of the Killing field takes longer to travel
than it would between the corresponding worldlines in Minkowski space\footnote{The spacetime is mapped
to Minkowski space by mapping each sphere of symmetry to one with corresponding area and preserving
angular relationships.}.

Because it refers to the time measured by an observer at infinity, this result has much of the
`asymptotic' flavor discussed above.  We therefore find it rather unsatisfying.  To be precise: if an
advanced civilization wished to send signals quickly in order to compete in some way against a
neighboring civilization, this theorem would provide useful guidance.  Assuming spherically symmetry and
a static spacetime, they should make their home as close to flat Minkowski space as possible.  But one
might imagine that a fast signaling time was desired for other reasons\footnote{After all, the density of
advanced civilizations in our galaxy appears to be rather small, so one might not expect competition to
be extreme.}. Perhaps it is desired ({\it e.g.}, for reasons of social coherence) to exchange signals on
a timescale that seems short to the participants involved; {\it i.e.}, as measured by the proper time
along the orbits $x,y$ of the Killing field?  One might imagine that organizations and individuals living
in distant parts of the civilization wish to exchange goods or information without receiving undue delays
(such as waiting a year to receive a much desired letter or package) due to limitations imposed by the
speed of light.

Let us therefore suggest the following two questions to provide a framework for our discussion.

\begin{quote}
{\bf Question 1: } Given a static region ${\cal V}$ of spacetime containing a sphere of
area $4\pi  R^2$, consider the proper time $2\tau_x$ along the orbit $x$ of the
Killing field required for a signal to propagate from $x$ to
another orbit $y$ on the sphere and then return.  Let $2\tau_{max}$ be the maximum such
signaling time between two such orbits.
What spacetime satisfying the dominant energy condition and
having such has a static region minimizes $\tau_{max}(R)$ and what is this
minimum?
\end{quote}
\begin{quote}

{\bf Question 2:} Consider a spherically symmetric asymptotically flat
spacetime of total mass $M$ and which is vacuum outside some sphere
of area $4 \pi R^2$.  Let us take $\tau_R = T /\sqrt{1 - 2M/R}$ to be
the Killing time normalized to measure proper time at the chosen
sphere.
What interior solution satisfying the dominant energy
condition allows a causal signal to
propagate between two given orbits $x$ and $y$ of
the time translation Killing field
in the smallest amount of time $\tau_R$, and what is this shortest
signaling time?
\end{quote}
In the case  that the infimum of $\tau_R,\tau_{max}$
is not realized by a smooth geometry, we take a sequence of smooth
geometries approaching the infimum to yield answers to these questions.
A number of related questions
also come to mind, but we will not discuss them here.

Question 1 is of intrinsic interest to an advanced civilization
wishing to create a `maximally convenient' home.  Question 2 is of interest
because it provides a setting in which, due to
Birkhoff's theorem\footnote{Which states that the spacetime outside
of the stated sphere will
necessarily the Schwarzschild spacetime with some mass $M$.} (see e.g. \cite{WaldBook}),
one feels confident that the sphere being discussed is in some sense
`the same sphere' regardless of how the interior is filled.  It also acknowledges the constraint
that, while the civilization may be able to modify there spacetime in the interior of their
sphere, they may have less control over the exterior region of the spacetime.

Note that in both cases we have required the dominant energy condition.
Recall that this
condition consists of the weak energy condition together with the requirement that $T^{\mu}_{\nu} t^{\nu}
$, if non-zero, should be a future directed timelike vector.
We choose it here because it is the strongest of the
usual energy conditions that is expected to hold for all reasonable forms of matter \cite{WaldBook},
unless one allows a negative cosmological constant.  We will have more to say about the
interplay between these questions and the choice of energy conditions in
section \ref{disc}.

The theorem of section \ref{bound} does provide some information
of interest to both questions.  For example, it gives a lower bound
on the signaling time as measured in both Question 1 (in the spherically symmetric setting)
and Question 2,
but it in no way guarantees that the bound can be saturated.
Because this information is incomplete, we are
motivated to explore the issue further by considering several particular spacetimes in detail
in section \ref{ex}.  While we are unable to answer either Question 1 or Question 2
in full, we identify features that may be of use in future investigations.  In particular,
we find that at least in regimes far from that containing a black hole one can come
close to saturating the bound of section \ref{bound}.  In addition,
it appears that the dominant energy condition is significantly more constraining than
is the weak energy condition.
These conclusions are discussed briefly in section \ref{disc}.

\section{A bound on fast spacetimes}
\label{bound}

In this section we prove a theorem showing that, as viewed from infinity,
static spherically symmetric spacetimes are never `faster' than Minkowski
space.  We begin with the following Lemma:

\begin{quote}
{\bf Lemma 1:} Consider an asymptotically flat spherically symmetric spacetime which is static for $r > r_0$, satisfies
$m(r_0)\ge 0$ as defined below,
and satisfies the {\it weak} energy condition.In such a spacetime, no clock with $r > r_0$ runs faster than a clock at
infinity.  That is, if the Killing time $T$ is normalized at infinity, the proper time $\tau$ of any
static clock increases no faster than $T$.
\end{quote}

Recall that spherically symmetric static metrics take the general form
 \bea \label{interior} d s^2=-f(r) d t^2 + h(r) d r^2 +
r^2d \theta + r^2 \sin^2\theta d \phi^2  \eea in Schwarzschild coordinates.
Our Lemma therefore is just
the statement that the metric function $f$ satisfies $f < 1$.
The metric function $h(r)$ is related to the spherically symmetric mass function $m(r)$ by
$h = (1-2m/r)^{-1}$.  Note that the above restriction on $m(r_0)$ is fulfilled whenever the region $r \ge r_0$
is part of a spacetime which satisfies the conditions of the positive mass theorem \cite{SY,Witten,AH,PT,Reula}.
  Note that we allow spacetimes with, {\it
e.g.} a central black hole with horizon at $r_0$.

The proof is straightforward.
Consider any spacetime satisfying the premises
stated above.
The metric components $g_{tt}=-f(r)$ and $g_{rr}=h(r)$ solve the Einstein equations with
sources given by
the stress-energy tensor $T^{\mu}_{\nu}=diag(-\rho,P_{r},P_{\theta},P_{\phi})$.
 These equations are equivalent to
\bea \label{first}
\partial_r m&=& 4 \pi r^2 \rho, \\
\label{second} \frac{\partial_r f}{2f} &=& \frac{m + 4 \pi r^3
P_{r}}{r(r- 2 m)}, \hspace{0.5cm} {\rm and}\\ \label{third}
 \partial_r P_{r} &=& -\left(\frac{\partial_r f}{2f}+\frac{2}{r}\right)
P_{r} - \frac{\partial_r f}{2f} \rho +\frac{2}{r} P_{\theta}. \eea
The last
of these (\ref{third}) encapsulates stress-energy conservation in
a spherically symmetric background.

Now consider the density profile $\rho_0(r)$  and the pressure profile $P_{r0}(r)$ in our spacetime. Since asymptotic flatness requires
$f=1$ at infinity, $f(r)$ is determined by integrating (\ref{second})
inward from infinity.  As a result, for a fixed density profile,
reducing the radial pressure at any $r$ increases $f$ at every smaller value of $r$.
Now recall that the given spacetime must satisfy the weak energy condition, which
requires $\rho_0 \ge - P_{r0}$.  Thus, if
we introduce a new spacetime
with the same density profile $\rho_0(r) $ but a new pressure profile $\tilde P_{r} = -\rho_0$
saturating the above bound,
the corresponding $\tilde f$ satisfies $\tilde f(r) > f(r)$ at each $r$.
Note that our new spacetime is described by the same function $h(r)$ as the original.
Now, since $h = (1 - 2m/r)^{-1}$, we have
\begin{equation}
\frac{\partial_r h}{2h} = \frac{-m + 4 \pi r^3 \rho}{r(r-2m)}.
\end{equation}
Since $\tilde P_r = - \rho_0$, comparison with (\ref{second}) shows that we have $\partial_r \ln h = -
\partial_r \ln \tilde f$; i.e., $\tilde f h = constant$. Evaluating this in the asymptotic region we find
$\tilde f = 1/h$.  But, using the timelike vector $\partial_t$ in (\ref{WEC}) yields $\rho \ge 0$ so that
 $m \ge 0$ from (\ref{first}) and $h =  (1 - 2 m/r)^{-1}  > 1$.  Thus $f < \tilde f < 1$, proving Lemma 1. In fact, our
result is somewhat stronger as we only used $\rho \ge -P_r$ (and not the entire weak energy condition).

With the aid of Lemma 1, it is now easy to prove the following theorem:
\begin{quote}
{\bf Theorem 1.} Consider a smooth, spherically symmetric, spacetime satisfying the {\it weak} energy
condition and static for $r >r_0$ and $m(r_0) \ge 0$.  Suppose the Killing time to be normalized at infinity and consider
two orbits $x$ and $y$ of the time translation symmetry lying on symmetry spheres with areas $4 \pi
R^2_x$ and $4 \pi R_y^2$ and separated by an angle $\theta$ on the spheres. Then, as viewed from
infinity, no signal staying within the static region can  be sent between $x$ and $y$ faster than one
could be sent if $x$ and $y$ lay on the corresponding sized spheres in Minkowski space with the same
angular separation; i.e., the Killing time $T$ required satisfies $T \geq \sqrt{R_x^2 + R^2_y -2 R_x R_y
\cos \theta}.$
\end{quote}

Using Lemma 1 and $h(r) >1$ we find that the signaling time satisfies

\begin{equation}
T = \int ds \sqrt{ \frac{h}{f} \dot{r}^2 (s) + \frac{r^2}{f} \dot{\theta}^2(s)} \ge \int ds \sqrt{
\dot{r}^2 (s) + r^2 \dot{\theta}^2(s) } \ge \sqrt{R_x^2 + R^2_y -2 R_x R_y \cos \theta}.
\end{equation}

Thus, we can in some sense show that Minkowski space is the `fastest' spherically symmetric static
spacetime.  However, this result has much of the `asymptotic' flavor that we wished to avoid.  In
particular, the notion of how `fast' the spacetime is has been referred to the observer at infinity.

Suppose we examine the implications of this theorem for Questions 1 and 2.
It shows that the signaling time between two orbits $x$ and $y$ on the sphere of area $4\pi R^2$ satisfies $T \ge 2R \sin \theta/2$ where $\theta$ is the angular separation of $x$ and $y$.
But, in terms of the proper time $\tau_R$ this is
$\tau_R \ge 2 R \left( \sin \theta/2 \right)
 \sqrt{1 - 2M/R}$.  From the perspective of observers on the shell this is a much weaker bound than they would find in Minkowski space.  It is therefore useful
to study the situation in more detail.  We begin this below by investigating a
number of examples.
While we will not succeed in identifying a `fastest'
spacetime,
we will learn much about the
problem, and find some interesting interaction with the
energy conditions.

\section{Some examples of `faster' spacetimes}

\label{ex}

We now proceed to explore Questions 1 and 2 in more depth through a number of examples.  We will make
frequent use of a certain strategy to explore the linearized change in the travel time near each of our
examples, so we present this method first in subsection \ref{lin}. We then study three families of
spacetimes in detail. All of these families satisfy the dominant energy condition, which is our primary
regime of interest.  The first family contains a Minkowski interior patched to the Schwarzschild exterior
via a thin shell. The other two correspond to various ways of  saturating of the energy conditions.
Although the last two cases will prove to be faster than the first one, perturbative analysis show that
there exist other spacetimes which are faster yet.

\subsection{Linearization Strategy}
\label{lin}

We will use the same notation for the metric and stress-energy as
in section \ref{bound}, though here we will be more concerned with
the dominant energy condition as required by Questions 1 and 2.
In our context this imposes
 \bea
\label{Energycond}
 \hspace{0.5cm} \rho \geq |P_{r}|, \hspace{0.5cm} \rho \geq
 |P_{\theta}|.
 \eea

For simplicity, we take the points $x$ and $y$ between which our signal is exchanged to lie
on the poles of the sphere at $r=R$. The fastest path connecting
them must be a null geodesic with $\phi=const$ so that the integral \bea \label{Time}
 T = \int ds \sqrt{\frac{h}{f} \dot{r}^2 (s) + \frac{r^2}{f}
 \dot{\theta}^2 (s)}
 \eea
provides the time of flight as measured by the Killing time $T$ normalized
at infinity.
We will also make use below of the change in (\ref{Time}) under small
perturbations of the metric (\ref{interior}).
The first order variation is
\bea \label{VarTimeone}
 \delta T = \int \frac{d s}{2 \sqrt{f}\sqrt{h \dot{r}^2+r^2
\dot{\theta}^2}} \left[ \dot{r}^2 \delta h - (h \dot{r}^2+ r^2
\dot{\theta}^2) \frac{\delta f}{f}  \right]. \eea However, a more
useful form is obtained by assuming a regular origin so that
boundary conditions imply $\delta m(0)= 0$. The
equations of motion (\ref{first})-(\ref{third}) then imply three
linear differential equations for the variations of $f$, $h$,
$\rho$, $P_r$ and $P_\theta$ which can be used to rewrite
(\ref{VarTimeone}) in terms of $\delta \rho$ and $\delta P_{r}$:
\bea
 \delta T&=& \int \frac{d s}{2 \sqrt{f}\sqrt{h \dot{r}^2+r^2
\dot{\theta}^2}} \left[ (2 h \dot{r}^2+ r^2 \dot{\theta}^2) \frac{\delta h}{h} - (h \dot{r}^2+ r^2
\dot{\theta}^2) \left( \frac{\delta h(R)}{h(R)} + \frac{\delta f(R)}{f(R)} \right)
+  \right.\nonumber \\
\label{VarTime} && \left.
 + 8 \pi (h \dot{r}^2+ r^2 \dot{\theta}^2) \int^{R}_{r} dr' r' [ h ( \delta \rho + \delta P_r)+
(\rho + P_r) \delta h ] \right], \eea where $\delta h= 2\frac{h^2}{r} \delta m = 8 \pi \frac{h^2}{r}
\int_0^r d r' r'^2 \delta \rho (r') $. We note for future reference that the derivation of
(\ref{VarTime}) uses the inner boundary condition (at the origin) but does not require any outer boundary
condition. All the examples we will study obey the equation of state $\rho + P_r = 0 $ in the relevant
region, so that, in order to maintain (\ref{Energycond}), perturbations should satisfy $\delta \rho +
\delta P_r \geq 0 $ . The existence of perturbations that generate negative $\delta T$ will prove that a
particular spacetime under study is not the fastest. To find such a perturbation,  we consider variations
having $\delta \rho + \delta P_r = 0 $ in order to eliminate the positive contribution of the last term
in (\ref{VarTime}).   In our applications below, we will also have $\delta f(R) =0$. Under these
assumptions, the expression for $\delta T$ reduces to \bea
 \delta
T&=& \int \frac{d s}{2 } \left(\frac{ 2 h \dot{r}^2+ r^2 \dot{\theta}^2}{\sqrt{f}\sqrt{h \dot{r}^2+r^2
\dot{\theta}^2}} \right) \frac{\delta h}{h} - \frac{1}{2} \int d s \sqrt{\frac{h}{f} \dot{r}^2 +
\frac{r^2}{f}
 \dot{\theta}^2}
 \frac{\delta h(R)}{h(R)} \nonumber \\
\label{VarTime2}
 &=& \int \frac{d s}{2 } \left(\frac{ 2 h \dot{r}^2+ r^2 \dot{\theta}^2}{\sqrt{f}\sqrt{h
\dot{r}^2+r^2 \dot{\theta}^2}}\right) \frac{\delta h}{h} - \frac{T}{2} \frac{\delta h(R)}{h(R)} . \eea

A spacetime with identically vanishing  $\delta T$ would be an excellent candidate for the fastest
spacetime.  While we have not been able to find such a solution consistent with the positive energy
condition, the result (\ref{VarTime}) is nevertheless quite useful in showing that the following simple
cases do not minimize the travel time (\ref{Time}).

\subsection{The empty shell spacetime}
\label{Mink}

Let us begin with the simplest allowed spacetime: a flat region inside the sphere of radius $R$ and, in
order to match a Schwarzschild exterior as required by Question 2, a thin shell at $r=R$ as determined by the discontinuity in
the extrinsic curvature across this surface. The metric of the empty interior $r < R$ is \bea
\label{emptyshell}
 ds^{2}=-(1-\mbox{$\frac{2M}{R}$}) dt^{2} +dr^{2}+r^{2}
d\Omega^{2}, \eea {\it i.e.},(\ref{interior}) with  $f= 1-2M/R$ and $h=1$. Here we have chosen the
normalization of $t$ so that $g_{tt}$ is continuous at $r=R$ as required by the Israel junction
conditions \cite{Israel}. The surface stress-energy of the shell must also satisfy the dominant energy
condition.

In general, the surface stress-energy tensor on a hypersurface
 $\Sigma$ is defined by the integral \cite{Israel}\bea
S^{\mu}_{\nu} =  \lim_{\epsilon \rightarrow 0}
\int^{+\epsilon}_{-\epsilon} T^{\mu}_{\nu} d n = \frac{1}{8 \pi}
\lim_{\epsilon \rightarrow 0} \int^{+\epsilon}_{-\epsilon}
G^{\mu}_{\nu} d n ,
 \eea
where $n$ is the proper distance measured along the normal to the hypersurface. For a general interior of
the form (\ref{interior}) and our Schwarzschild exterior, this surface tensor is
\bea \label{surf}
 8\pi S^{t}_{t}= -\frac{2}{R}
\left( \sqrt{1-\mbox{$\frac{2M}{R}$}}-\frac{1}{\sqrt{h}} \right), \hspace{0.3cm} S^{r}_{r}=
0,\hspace{0.3cm}
 8 \pi S^{\theta}_{\theta} =
\frac{1-\frac{M}{R}}{R \sqrt{1-2\frac{M}{R}}}-\frac{1}{\sqrt{h}}\left( \frac{f'}{2f}+\frac{1}{R} \right),
\eea where the metric components and their derivatives are evaluated by taking the limit $r \rightarrow
R$ from below.  We will continue to use this convention: any discontinuous function evaluated at $R$ is
to be understood as the limit $r \rightarrow R$ from below.

In particular we find \bea \label{emptySurf} 8\pi S^{t}_{t}= \frac{-2}{R} \left(
\sqrt{1-\mbox{$\frac{2M}{R}$}}-1 \right),\hspace{0.5cm} 8 \pi S^{\theta}_{\theta} = \frac{1}{R} \left(
\frac{1-\frac{M}{R}}{\sqrt{1-2\frac{M}{R}}}-1 \right) \eea for the empty shell metric (\ref{emptyshell}).
One may then check that
 $S^{t}_{t} \geq |S^{\theta}_{\theta}|$ is satisfied exactly in the range
$0 \leq \frac{M}{R} \leq \frac{12}{25}$.

It is clear that the fastest trajectory follows a radial path with $ \dot{\theta}= 0$, so that, from
(\ref{Time}), the travel time is
\bea T_{Empty}=\frac{2R}{ \sqrt{1-\frac{2M}{R}}}. \eea In terms of the
proper time $\tau_R$ measured by a static observer at $r=R$ this is just $\tau_R^{Empty} = 2R$.

Although this spacetime is a natural one to study, it is not the fastest.  This may be seen by
considering the variation (\ref{VarTime2}) of $\delta T$ under a perturbation $ \delta \rho (r)=- \delta
P_{r} (r)=-\delta P_{\theta}(r)=\delta \rho (0) >0$, so that $\delta h(r) =\frac {8 \pi}{3} \delta \rho
(0) r^2 $.  Since the signal takes no time to cross the shell, it is sufficient to apply (\ref{VarTime2})
at some $r$ just a bit less than $R$. We will not need the explicit form of the perturbation at the shell
since, due to the continuity of $f$ at $r=R$ we have $\delta f(R)=0$ so we may use equation
(\ref{VarTime2}) for the variation $\delta T$.  The perturbed spacetime clearly satisfies the energy
conditions in the interior and, since the original shell at $r=R$ does not saturate these
conditions\footnote{The case with $\frac{M}{R}=\frac{12}{25}$ does in fact saturate $S_t^t\geq
|S^{\theta}_{\theta}|$  and requires more care. It may be treated as in section \ref{ds} below.}, there
is no danger that they will be violated at $r=R$ for small $\delta \rho$. For this perturbation one finds
$\delta T = -\frac{8 \pi \delta \rho(0) R^{3}}{9 \sqrt{1-\frac{2 M}{R}}}<0$ for a radial trajectory, so
that the empty shell spacetime is not the fastest.

\subsection{De Sitter space in a bottle}
\label{ds}

Since the only constraints in our problem are the energy
conditions, one might expect these conditions to be saturated by
our hypothetical fastest spacetime. The weak energy condition is
saturated by taking $\rho=-P_{\theta}=-P_{r} >0$, in which case
stress-energy conservation requires $\rho(r)$ to be just some
constant $\rho_0$. In the previous subsection we found
the travel time to be reduced by perturbing our empty shell
spacetime in this direction. Unfortunately, such a spacetime does
match the boundary condition that $\rho =0$ for $r> R$ as required by Question 2.

On the other hand, this discussion suggests that one might study
the spacetime we call ``De Sitter space in a bottle'' in which we
take $\rho=-P_{\theta}=-P_{r} = \rho_0 >0$ for $r < R$ but
add a shell at $r=R$ to satisfy stress-energy conservation. This
shell effectively constitutes a `bottle' whose stresses and
gravitational self-attraction keeps the piece of de Sitter space
with $r< R$ from expanding.

The metric takes the form \bea \label{deSitter}
ds^{2}=-(1-\mbox{$\frac{2M}{R}$}) \frac{1-b^{2}
\frac{r^{2}}{R^{2}}}{1-b^{2}}dt^{2} + \frac{1}{1-b^{2}
\frac{r^{2}}{R^{2}}} dr^{2} + r^{2} d\Omega ^{2}, \eea where $b^2
=  \frac{8 \pi}{3} \rho_{0} R^2  <1$. Here, $t$ has again been
normalized in the interior so that $g_{tt}$ is continuous across
$r=R$.  Comparing (\ref{deSitter}) and (\ref{surf}), one finds the
surface stresses to be \bea
 \label{dSSurf}
8 \pi S^{t}_{t}= \frac{-2}{R} \left(\sqrt{1-\mbox{$\frac{2M}{R}$}}-\sqrt{1-b^{2}}\right), \hspace{0.5cm}
 8 \pi S^{\theta}_{\theta} = \frac{1}{R}
\left( \frac{1-\frac{M}{R}}{\sqrt{1-\frac{2M}{R}}}-\frac{1-2 b^2}{\sqrt{1-b^2}} \right). \eea Imposing
the dominant energy condition at the shell requires \bea \label{bminus} b^2 \leq
b^{2}_-(\mbox{$\frac{M}{R}$}) = \frac{3}{4}- \frac{S(\frac{M}{R})}{32}-\frac{S(\frac{M}{R})}{32}
\sqrt{1+\frac{16}{S(\frac{M}{R})} }, \hspace{0.3cm} {\rm where} \hspace{0.3cm}  S( \mbox{$\frac{M}{R}$} )
= \frac{(3-5 \frac{M}{R})^{2}}{1-2 \frac{M}{R}}, \eea  and $ 0 \leq \frac{M}{R} \leq \frac{12}{25}$, as
in the previous example\footnote{For configurations satisfying $\rho=-P_r$ in the interior, the condition
$S_t^t \geq |S^{\theta}_{\theta}|$ can be rewritten as $\sqrt{S(\mbox{$\frac{m(R)}{R}$}
)}-\sqrt{S(\mbox{$\frac{M}{R}$} )} \geq \frac{ 4 \pi R^2 \rho(R)}{\sqrt{1-\frac{2m(R)}{R}}} \geq 0$. In
order to have $M \geq m(R)$, the form of $S$ requires $ 0 \leq \frac{M}{R}\leq \frac{12}{25}$.}.

As for the empty shell spacetime, the antipodal orbits $x$ and $y$ can be connected only by radial
geodesics\footnote{Note that a non-radial geodesic would lead to an $S^1$ of such geodesics, and thus to
a light cone with a caustic at finite affine parameter. As it is readily seen from the conformal diagram
(see, {\it e.g.}, \cite{HawEllis}), this does not occur in de Sitter space.}. Thus, we again have
$\dot{\theta}=0$ and the travel time is \bea \label{TdeSitter}
\tau_R&=&T(b) \sqrt{1- \frac{2M}{R}} \nonumber \\
 &=&\tau_R^{Empty} \frac{\sqrt{1-b^{2}}}{2b} \ln{
\frac{1+b}{1-b}} .
\eea
 Since the factor multiplying $\tau_R^{Empty}$ is
less than $1$, this space is faster than the empty shell spacetime. As (\ref{TdeSitter}) is a
monotonically decreasing function of $b>0$, the smallest allowed time (for fixed $\frac{M}{R}$) occurs
when (\ref{bminus}) is saturated, {\it i.e.}, when $S^{t}_{t} = |S^{\theta}_{\theta}|$. In particular,
the largest effect occurs for $\frac{M}{R}=\frac{2}{5}$, when (\ref{bminus}) gives the biggest allowed
$b$, $b=\frac{19}{32}\left( 1-\sqrt{\frac{105}{361}} \right)$, and, therefore, the smallest value of
(\ref{TdeSitter}), $\frac{\tau_R}{\tau_R^{Empty}} \approx 0.987$.

However, a perturbation analysis again shows that
 spacetimes outside this class are faster yet.  Again we apply
(\ref{VarTime2}) to the region inside the shell. Let us denote the perturbed quantities with tildes.  A
perturbation $ \delta \rho (r)=- \delta P_{r} (r)=-\delta P_{\theta}(r)=\delta \rho (0)>0$, corresponding
to $\tb = b_- + \delta b > b_- $ would reduce the time by the amount $\delta_0 T \equiv T(\tb)-T(b_-)<
0$, but, it would also violate the energy condition $\tS^{t}_{t} \geq |\tS^{\theta}_{\theta}|$. In order
to respect the dominant energy condition at the shell, we instead use a sequence of perturbations
$\{\delta_n \rho \}$ of the form \bea \label{rhopert} \delta_n \rho(r) &=& -\delta_n P_r(r)=A_n (r-r_n)+
\delta \rho (0) \ \ \ {\rm for} \ \ \ R
> r>r_n, \cr  \delta_n \rho(r) &=& -\delta_n P_r(r)= \delta \rho (0) \ \ \  {\rm for} \ \ \ r<r_n. \eea which
differ from $\delta \rho (0)$ in the region $R > r > r_n$. Our goal will be to satisfy the energy
conditions for large enough $n$. Here $A_n$ are constants and linearized stress-energy conservation
together with the energy condition in the interior requires $A_n = - \frac{2}{r} (\delta_n \rho +
\delta_n P_{\theta}) < 0$. Similarly, at the shell the dominant energy condition requires \be
\label{DECS2} \tS^t_t-\tS^{\theta}_{\theta}=\delta_n S^t_t- \delta_n
S^{\theta}_{\theta}=\frac{1}{R^2\sqrt{1-b_-^2}}\left[-4 \pi R^3 \delta_n \rho (R)
-\frac{2-b_-^2}{1-b_-^2} \delta_n m (R)\right]  > 0. \ee

Let us choose $r_n$ to converge to $R$ and also require each $\delta_n \rho $ to yield the same value
$\delta m (R)$ for the change in the mass function $m(r)$ evaluated just inside the shell. Note that this
is readily achieved by taking $A_n$ to scale with $(R-r_n)^{-2}$. In this case (\ref{DECS2}) is indeed satisfied for
sufficiently large $n$.

It is clear that at each point $r$ in the interior $\delta_n \rho(r)$ converges to $\delta \rho(0)$.
Thus, it makes sense to express the variation $\delta_n T$ of $T$ under $\delta_n \rho$ in terms of the
variation $\delta_0 T$ obtained by the constant density perturbation associated with simply shifting $b$.
{}From (\ref{VarTime2}) we find in the limit
\be \label{TpertdS} \delta_n T \rightarrow \delta_0 T+
\frac{\delta_0 m(R)-\delta m(R)}{\sqrt{1-\frac{2R}{M}}} \frac{1}{b\sqrt{1-b^2}} \ln{\frac{1+b}{1-b}},
\ee where on $\delta_0 m (R)$ refers to the change in the mass $m(r)$ evaluated just inside the shell
under the constant density perturbation associated with changing the density uniformly by $\delta \rho
(0)$.

Since $A$ is negative, the second term in (\ref{TpertdS}) is positive. However, it is clear from the
construction of $\delta_n \rho$ that we are free to take $\delta m(R)$ as close as desired to $\delta_0
m(R)$ without changing $\delta b$.  As a result, this second (positive) term can be made negligible in
comparison with the first (negative) term.  We have therefore established the existence of small
perturbations which preserve the positive energy conditions but reduce the travel time below that of the
background ``dS in a bottle'' spacetime. A similar analysis applies to the empty shell spacetime in the
extreme cases $\frac{M}{R} = \frac{12}{25}$.

Because we imposed the dominant energy condition, spacetimes in this class were restricted to be much
slower than would be guaranteed by Theorem 1. In contrast, note that we can do much better if we enforce
only the weak energy condition. This will require $S_t^t \ge 0$ and thus $b^2 \le 2M/R$, but this is the
only requirement.  Note that this is just the condition that $m(R) \leq M$; {\it i.e.}, that the mass
contained in the region $r < R$ is less than or equal to the total mass $M$ of the spacetime. Denoting
the bound set by Theorem 1 by $\tau_R^{bound}$ and comparing with (\ref{TdeSitter}) for $b^2 = 2M/R$, one
finds

\begin{equation}
\frac{\tau_R}{\tau_R^{bound}} = \frac{1}{2b} \ln \frac{1+b}{1-b}.
\end{equation}
So, for $b^2 = 2M/R \sim 1$, we find $ \tau_R \gg \tau_R^{bound}$. Nevertheless, $\tau_R \rightarrow 0$
so that $\tau_R \ll \tau_R^{Empty}$. Thus, this example suggests that the dominant energy condition may
be significantly more restrictive that the weak energy condition in investigating Questions 1 and 2.

\subsection{Saturating the dominant energy condition}
\label{last}

We now turn to our third example. We saw in the proof of Lemma 1 that it was advantageous to set $\rho=
-P_r$ and take $\rho$ as large as possible. The same is true with our current boundary conditions.
However, stress-energy conservation places bounds on how rapidly $P_r$ may change.  In particular, we can
rewrite (\ref{third}) as \bea \label{sEc}
\partial_r P_{r}= - \frac{\partial_r f}{2f}(\rho + P_{r}) +
\frac{2}{r}(\rho - P_{r})-\frac{2}{r}(\rho-P_{\theta}).
 \eea
Maintaining $P_r = -\rho$ with a rapidly changing
$\rho(r)$ may force $P_\theta$ to be very large and perhaps to
violate the dominant energy condition $P_\theta <
\rho$. In fact, if one has already imposed $P_r = -\rho$, taking
$P_\theta = \rho$ allows $P_r$ to decrease at the fastest possible
rate as one moves away from the boundary.

As a result, we are motivated to consider spacetimes with $\rho=-P_{r}= P_{\theta}$.  Stress-energy
conservation (\ref{sEc}) then requires \bea \rho(r)=\rho_{0}(\frac{r_{0}}{r})^{4}, \eea for constants
$\rho_0 \ge 0$ and  $0< r_{0} < R$. To evade the divergence at $r=0$, we excise the region $r < r_0$ and
sew in a piece of another spacetime.  For lack of a better choice, we once again use a piece of de Sitter
space. We demand that $\rho$ is continuous at $r_ {0}$ so that $m$ is $C^1$ and there is no additional
shell of mass at this junction.

The mass function is \bea \label{dSmassone} m = \Biggl\{ { \hspace{2.7cm} {\frac{4 \pi}{3} \rho_{0} r^{3}
\hspace{2.2cm} {\rm for} \ \ \ 0 \leq r\leq r_{0}} \atop {- 4 \pi \rho_{0} r^{4}_{0}\left( \frac{1}{r}
\right) +\frac{16 \pi}{3} \rho_{0} r^{3}_{0} = -\frac{c}{r}+a \ \ \ {\rm for} \ \ \ r_{0} \leq r<R.}}
\eea where $ r_{0}=\frac{4 c}{3a}$ and $ \frac{ 4 \pi}{3} \rho_{0}= (\frac{3}{c})^{3} (\frac{a}{4})^{4}$.
We refer to this case as the  ``dS/DEC" spacetime due to the saturation of the dominant energy condition
for $r > r_0$ and the presence of the de Sitter region for $r< r_0$.

 Let us introduce the dimensionless
variables \bea
 \hatt &=&\frac{2 t}{T_{Empty}}= \frac{t}{R} \sqrt{1-
 \mbox{$\frac{2M}{R}$}},\hspace{0.5cm} \hr = \frac{r}{R}, \hspace{0.5cm} \ha
 =\frac{a}{R},\hspace{0.5cm} \hc = \frac{c}{R^{2}}, \hspace{0.5cm} \hm(\hr)
 =\frac{m(r)}{R},
 \eea
in terms of which the metric takes the form
\bea
 \label{satDEC}
ds^{2}= R^{2}\left[-\frac{1-\frac{2 \hm(\hr)}{\hr}}{1-2 \ha+ 2 \hc} d \hatt^{2} + \frac{1}{1-\frac{2
\hm(\hr)}{\hr}} d \hr^{2} + \hr^{2} d \Omega^{2} \right]. \eea Note that for $\hc = \frac{3 \ha}{4}$, the
de Sitter region fills all of $r< R$. As a result, we require $\hc \leq \frac{3 \ha}{4}$. The value $
\hr_{BH}= \ha + \sqrt{ \ha^{2}- 4 \hc}$, which is real for $ \hc \leq \frac{\ha^{2}}{2}$, would
correspond to the location of a Killing horizon,{\it i.e.}, $g_{tt}(\hr_{BH})\propto 1-\frac{2 \ha}{
\hr_{BH}} + \frac{2 \hc}{\hr_{BH}^{2}} =0$. But note that $ \hc \leq \frac{\ha^{2}}{2}$ yields $ \hr_0
\leq \frac{2 \ha}{3} < \ha < \hr_{BH}$. Thus, to avoid the existence of a horizon\footnote{One could
consider spacetimes with a black hole instead of a dS interior, but then there are no radial null
geodesics connecting antipodal points on the sphere.  We explored the behavior of selected non-radial
null geodesics numerically in such a spacetime but in each case found $T > T_{Empty}$.  For this reason
we chose to concentrate on radial geodesics and on spacetimes that allow them.}, we must have $ \hc >
\frac{\ha^{2}}{2}$.

We also investigate any further restriction imposed by
requiring the shell to satisfy the dominant energy condition.
Again using (\ref{surf}), the relevant stresses are
\bea
 \label{DECSurf} \hspace{-0.4cm}
8 \pi S^{t}_{t}= \frac{-2}{R} \left( \sqrt{1-\mbox{$\frac{2M}{R}$}}-\sqrt{1-2 \ha + 2 \hc} \right), 8 \pi
S^{\theta}_{\theta} = \frac{1}{R} \left( \frac{1-\frac{M}{R}}{\sqrt{1-\frac{2M}{R}}}-\frac{1- \ha
}{\sqrt{1-2 \ha + 2 \hc }} \right). \eea Condition $S^{t}_{t} \geq |S^{\theta}_{\theta}|$ also
constraints the values of $(\ha,\hc)$ through \be \label{cplus}
 \hc \geq -\frac{3}{4}+\frac{5}{4} \ha + \frac{1}{16}
S(\mbox{$\frac{M}{R}$}) +\frac{1}{16} S(\mbox{$\frac{M}{R}$})
\sqrt{1+\frac{8(\ha-1)}{ S(\frac{M}{R})}},
 \ee where $ S(\frac{M}{R})$ is again
as in (\ref{bminus}) and $ 0 \leq
\frac{M}{R} \leq \frac{12}{25}$. A plot of the allowed regions in the $\ha \hc$ plane for three different
values of $\frac{M}{R}$  is shown in figure 1. Curves of the form
(\ref{cplus}) move to the right in the $\ha \hc$
plane for increasing $ \frac{M}{R} \leq \frac{2}{5}$, and back to the left for $\frac{M}{R} >
\frac{2}{5}$.

\begin{figure}
\label{fig1} \centering
\includegraphics[width=8cm]{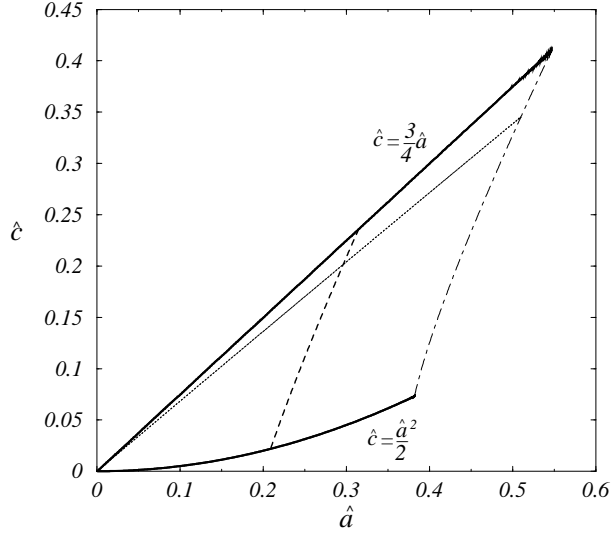}
\caption{The allowed region in the $\ha\hc$ plane for configurations (\ref{satDEC}) is given by
 $ \frac{ \ha^2}{2} < \hc \leq \frac{3 \ha}{4}$ and  condition (\ref{cplus}). The case $\hc=\frac{3\ha}{4}$
 represents the  ``dS in a bottle"  spacetime of section \ref{ds}.
 The dash-dotted line is obtained by setting $ \frac{M}{R}=\frac{2}{5}$ in equation (\ref{cplus}). For other values of $\frac{M}{R}$,
 the allowed region becomes smaller, as shown by the dashed line which
 represents condition (\ref{cplus}) for both $\frac{M}{R}=\frac{1}{5}<\frac{2}{5}$ and $\frac{M}{R}=\frac{7}{15}>\frac{2}{5}$.
 The thin dotted line indicates the points $ (\hc_{min},\ha_{min}) $ where the time (\ref{TsatDEC}) attains its minimum values in the allowed regions for each $\frac{M}{R}$.
}
\end{figure}

For a radial trajectory, we can explicitly write down the
expression for the time of flight
\begin{eqnarray} \nonumber
 \hT(\ha,\hc) &=& \frac{T(\hat{a},\hat{c})}{T_{Empty}}
 \\&=& \sqrt{1-2 \ha + 2 \hc}   \left[ \frac{4}{\ha} \left(\frac{2 \hc}{3 \ha}\right)^{\frac{3}{2}}\ln{\left(
     \frac{1+\sqrt{\frac{3 \ha^2}{8 \hc}}}{1-\sqrt{\frac{3 \ha^2}{8 \hc}}}\right)}
     + 1-\frac{4 \hc}{3 \ha} + \ha \ln{\frac{1-2 \ha +
2\hc}{(\frac{4 \hc}{3 \ha})^2(1-\frac{3 \ha^2}{8 \hc})}} - \right. \nonumber \\
\label{TsatDEC} && \left. -\frac{\sqrt{2\hc}(1-\frac{\ha^2}{\hc})}{\sqrt{1-\frac{\ha^2} {2 \hc}}}  \left(
\arctan {\frac{1-\ha}{\sqrt{2 \hc}\sqrt{1-\frac{\ha^{2}}{2\hc}}}}-\arctan { \frac{\frac{4 \hc}{3
\ha}(1-\frac{3\ha^{2}}{4 \hc})}{\sqrt{2\hc}\sqrt{1-\frac{\ha^{2}}{2\hc}}}} \right) \right].
\end{eqnarray}
This is complicated to study analytically. We have therefore used a simple C++ program to compute the
minimum value of $\hT$ for each $\frac{M}{R}$. The results are plotted in figure 2 and show a minimum at
$M/R = 2/5$ at a value of approximately $0.939$.

Note that
$\hT_{min}(\frac{M}{R})$ decreases monotonically for $ 0< \frac{M}{R}< \frac{2}{5}$.
Since, in this interval, the allowed region of parameters $(\ha,\hc)$ grows monotonically with $ \frac{M}{R} $, the minimum of $\hT$ for each $ \frac{M}{R} $
must be attained on the boundary of the allowed region.  This means that the minimum
occurs where $S^{t}_{t} \geq |S^{\theta}_{\theta}|$ is
saturated. A similar analysis applies for $ \frac{2}{5}< \frac{M}{R}< \frac{12}{25} $.

Note that the uppermost curve ($\hc=\frac{3\ha}{4}$) in figure 1 represents the ``dS in a bottle''
spacetimes.  Since it does not cross the middle curve showing the fastest ``dS/DEC'' spacetimes, we see
that ``dS in a bottle'' is never the fastest case and we have indeed improved upon the results of section
\ref{ds}.

\begin{figure}
\centering
\includegraphics[width=8cm]{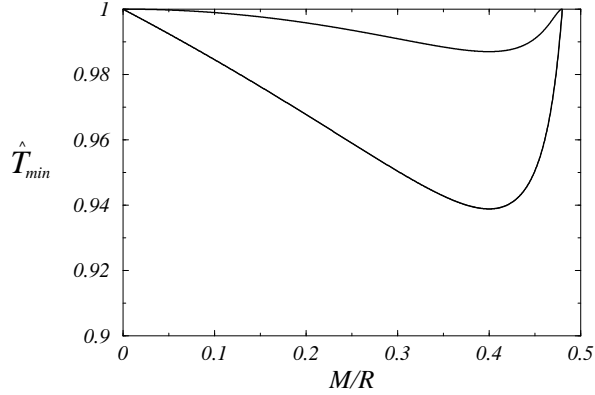}
 \caption{ The minimum time of flight (\ref{TsatDEC}) in ``dS/DEC" configurations as a function of
 $(\frac{M}{R})$ is represented by the lower curve. For comparison, the upper curve shows the minimum time of
 flight in the (slower)
``dS in a bottle'' configurations as a function of $ (\frac{M}{R})$. }
\end{figure}

Perturbing around configurations $(c_{min},a_{min})$ once again shows that the signaling time for this
family of spacetimes  can be reduced by perturbations outside of the family. Let us begin with the
observation that we have already shown that the time of flight would decrease if we were allowed to move
farther to the right in figure 1 for the same $M,R$. This corresponds to a perturbation $\delta_0 \rho $
satisfying $\delta_0 \rho + \delta_0 P_r=0$ in the interior and preserving the dominant energy condition
in the interior. However, it leads to a violation of the dominant energy condition at the shell.  We
therefore follow the strategy used in section \ref{ds} of adapting this initial guess (which we call
$\delta_0 \rho, \delta_0 m, \delta_0T$) to form a sequence of perturbations ($\delta_n \rho, \delta_n m,
\delta_n T$) which preserve the dominant energy condition at the shell for large enough $n$.

This condition requires:
\be
\label{deltaS} \tS^t_t-\tS^{\theta}_{\theta}=\delta S^t_t- \delta
S^{\theta}_{\theta}=\frac{1}{R^2\sqrt{1-2a+2c}}\left[- R \,
\delta m' (R) -\frac{2-5a+6c}{1-2a+2c} \delta m
(R) \right]
> 0.
\ee

Each perturbation $\delta_n \rho$  will be associated with a radius $r_n$ such that $\delta_n \rho =
\delta_0 \rho > 0$ for $r < r_n$.  We take the $r_n$ to increase with $n$ and to converge to $R$.  Choose
some $r_1$ and let $\delta_1 \rho$ be any such smooth perturbation which decreases for $r_1 < r < R$.
Such a $\delta_1 \rho$ will respect the dominant energy condition in the interior.  For later use, we
also require that the induced change $\delta_1 m(R)$ in the mass function just below the shell satisfy
$\delta_1 m(R) < \delta_0 m(R)$.

We now take $ \delta_n \rho$ to induce the same change in the mass just inside the shell for all $n$:
$\delta_n m(R) = \delta_1 m(R)$. We also require $\delta_n \rho$ to be a decreasing function of $r$, and
the sequence $\{ \delta_n \rho\}$ to have the property that $\delta_n \rho=\frac{\delta_n m'}{4 \pi r^2}$
become large and negative at $r=R$ when $n$ becomes large and $r_n \rightarrow R$. Then (\ref{deltaS})
is clearly satisfied for large $n$.

Since on the other hand $\delta_n \rho (r) \rightarrow \delta_0 \rho(r)$ for $r < R$, we find
\be
\label{Tpertdec} \delta_n T \rightarrow \delta_0T+ [\delta_0 m(R)-\delta_1 m(R)] \frac{T}{R-2a+ 2c/R }.
\ee
As in section \ref{ds}, the first term is negative by construction, and the second term can be chosen
to be arbitrarily small.  Thus, we have demonstrated the existence of perturbations of the ``dS/DEC''
spacetime preserving the dominant energy condition and further reducing the signaling time between
antipodal points.

\section{Discussion}

\label{disc}

In this work we have investigated the possibility of fast travel in static spherically symmetric
spacetimes.  We derived a simple theorem to the effect that, when the signaling time is measured by an
observer at infinity, a signal propagating through a spacetime satisfying the (timelike) weak energy
condition never arrives at its destination sooner than would a corresponding signal in Minkowski space.
This may be considered a non-perturbative generalization of \cite{VBL}.  Spherical symmetry and the
static Killing field were essential in identifying a corresponding signal in Minkowski space.

However, we were not satisfied with this result and wished to investigate related questions
concerning more local notions of signaling time.  For example, it is of interest whether
the observers who send and receive the signals find the propagation time to be less or
greater than what they would expect based on their Minkowski space intuition.
The theorem of section \ref{bound} does place a lower bound on this signaling time, but
it is a bound that is arbitrarily small compared to the naive Minkowski signaling time\footnote{i.e., a proper time of $2R$ for a light ray to propagate across a sphere of area $4\pi R^2$.} when
the signal propagates near the horizon of a black hole.  We also wished to explore
the consequences of requiring stronger energy conditions to hold.

For this reason we investigated several families of spacetimes in detail.  We were most interested in
cases where the {\it dominant} energy condition holds.   With this restriction, we found that we could
indeed construct positive energy spacetimes that improve upon the naive Minkowski time of $2R$, but only
by factors of order one. Our fastest such spacetime improves this result by approximately 6\%,
Perturbative analysis tell us that spacetimes exist which are faster yet, but of course give us no idea
of how much faster they might be.  There thus remains a sizable gap\footnote{When $\frac{2M}{R} \sim 1$.
On the other hand, for $\frac{2M}{R} \ll 1$ the bound is of course close to the naive Minkowski estimate:
$\tau_R^{bound} = 2R (1 + O(M/R))$.} between the fastest spacetime known to us and the bound we have
derived.  Discovering how this gap may be closed remains an open issue for future research, as does the
exploration of other variants of Questions 1 and 2.

Perhaps the most interesting suggestion from our investigation is that imposing only the weak energy
condition may allow much faster spacetimes. In particular, we found in section \ref{ds} that we could
construct spacetimes satisfying the weak energy condition which allowed signaling across our sphere in a
proper time significantly faster than $2R$.   For $2M/R \sim 1$ we found that while our signaling time
was much larger than the bound of Theorem 1, it could be made arbitrarily short compared to the naive
Minkowski bound.

Most of the work to date has considered the (null) weak energy condition because it leads
to powerful analysis techniques based on the Raychaudhuri equation and focussing theorems.
In our case, we saw that the weak energy condition led directly to our lemma and our
theorem in section \ref{bound}.   One would expect that both of these results to
generalize beyond the spherically symmetric context and to again require only the
weak energy condition for their proof.

On the other hand, realistic spacetimes should also satisfy the dominant energy condition\footnote{Unless one allows a negative
cosmological constant.}.  Thus, our examples suggest that they should be subject
to significantly stronger constraints.
If this is indeed the case, new analysis tools more appropriate to the
dominant energy condition will
need to be constructed before one can conclusively identify the fastest DEC spacetime
and the fastest allowed signaling time. We leave this task for future work.

%%%%%%%%%%%%%%%%%%%%%%%%%%%%%%%%%%%%%%%%%%%%%

\begin{acknowledgments}

D.M. would like to thank Bob Wald and Rafael Sorkin for a very
interesting conversation during which the idea for this project
first arose. Thanks also to   Antonio de Felice, Jim Hartle, and Rafael Sorkin for a variety of comments and questions.
Finally, we thank Joel Rozowsky for help with the figures and Harvey Reall for pointing out an error in an earlier version of this
paper.
This work was supported by in part by
NSF grants PHY97-22362 and PHY00-98747, the Alfred P. Sloan
foundation, and by funds from Syracuse University.

\end{acknowledgments}

%%%%%%%%%%%%%%%%%%%%%%%%%%%%%%%%

%%%%%%%%%%%%%%%%%%%%%%%%%%%%%%%

\end{document}